\documentstyle[11pt]{article}
\begin{document}
\title{Transition Matrix Monte Carlo Method\footnote{Talk presented
on CCP 1998, 2--5 September 1998, Granada, Spain.}}

\author{Jian-Sheng Wang\\
Department of Computational Science,\\
National University of Singapore,
Singapore 119260}

\date{20 October 1998}

\maketitle

\begin{abstract}

We analyze a new Monte Carlo method which uses transition matrix in
the space of energy.  This method gives an efficient reweighting
technique.  The associated artificial dynamics is a constrained random walk in
energy, producing the result that correlation time is proportional to
the specific heat.
\end{abstract}

One of the important application of Monte Carlo method
\cite{Kalos,Binder} is to compute very high-dimensional integrals
which appear in statistical mechanics.
The method generates a sequence of states $X_0$, $X_1$, $X_2$, \dots,
with a transition probability $T(X\to X') = P(X'|X)$.  If we want that
the distribution of $X$ follows $P_{eq}(X)$, it is sufficient to
require that
\begin{equation}
P_{eq}(X)\, T(X\to X') = P_{eq}(X')\, T(X' \to X).
\end{equation}
This is known as detailed balance condition.

We consider a simple classical spin model, the Ising model, as an
example of Monte Carlo dynamics.  The spins take only two possible
values and live on the sites of a lattice, for example on a square
lattice.  The total energy is a sum of interactions between
nearest-neighbor sites.  In a single-spin-flip dynamics, a Monte Carlo
move consists of picking a site at random, and flipping the spin with
probability \cite{Glauber}
\begin{equation}
 w = { 1\over 2} \left[ 1 - \sigma_0 \tanh\left( {J\over k_B T} \sum_i 
\sigma_i \right) \right], \label{weq}
\end{equation}
where $\sigma_0$ is the spin value before the flip, and $\sum_i
\sigma_i$ is the sum of spins of the nearest neighbors.  Another
popular choice is the Metropolis rate $\min\bigl(1, \exp(-\Delta H/k_B
T) \bigr)$ where $\Delta H$ is the energy increase due to flip.

The local Monte Carlo dynamics has some common features: (1) The
algorithm is extremely general.  It can be applied to any classical
model.  (2) Each move involves $O(1)$ operations and $O(1)$ degrees of
freedom.  (3) The dynamics becomes slow near a critical point,
characterized by divergent time scale,
\begin{equation}
   \tau \propto L^z, \qquad T = T_c, \qquad z \approx 2,
\end{equation}
where $L$ is system linear size.  See ref.~\cite{Wang-Gan} and
references therein for recent results on $z$.

Cluster algorithms \cite{SW} have very different dynamical
characteristics.  The Swendsen-Wang
algorithm uses a mapping from Ising model to a
type of percolation model.  Each Monte Carlo step consists of putting
a bond with probability
\begin{equation}
    p(\sigma_i, \sigma_j) = 1 - 
        \exp\bigl( -J(\sigma_i \sigma_j + 1)/k_B T\bigr)
\end{equation}
between each pair of nearest neighbors.  by ignoring the spins and
looking only at the bonds, we obtain a percolation configurations of
bonds \cite{Stauffer}.  A new spin configuration is obtained by
assigning to each cluster, including isolated sites, a random sign
$+1$ or $-1$ with equal probability.

Some of the salient features of cluster algorithms: (1) The algorithm
is applicable to models containing Ising/Potts symmetry. (2) Computational
complexity is still of $O(1)$ per spin per Monte Carlo step.  (3) Much
reduced critical slowing down.  The dynamical critical exponent $z_{sw}$ 
in $\tau \propto L^{z_{sw}}$ is roughly 0, 0.3, 0.5, 1,
in dimensions 1, 2, 3, and 4 or higher, respectively.  In
addition, Li and Sokal \cite{Li-Sokal} showed that $\tau \ge a c$ for
some constant $a$, and $c$ is the specific heat.

The transition matrix Monte Carlo \cite{Swendsen-Li} is related to the
single-spin-flip dynamics in the following sense, but it has a totally
different dynamics.  A single-spin-flip Glauber dynamics of the Ising
model is described by
\begin{eqnarray}
{\partial P(\sigma, t) \over \partial t} & =  &
\sum_{\{\sigma'\}} \Gamma(\sigma,\sigma') P(\sigma', t)  \nonumber \\
& = & \sum_{i=1}^N \Bigl[ -w_i(\sigma_i) + w_i(-\sigma_i)F_i\Bigr] P(\sigma,t),
\label{Meq} 
\end{eqnarray}
where $N$ is the total number of spins, and $w$ is given by
Eq.~(\ref{weq}), and $F_i$ is a flip operator such that $F_i
P(\ldots,\sigma_i, \ldots) = P(\ldots,-\sigma_i, \ldots)$.  Transition
matrix Monte Carlo dynamics is {\sl defined} by
\begin{equation}
 { \partial P(E,t) \over \partial t }  = \sum_{E'} T(E,E') P(E',t), \label{Teq}
\end{equation} 
where $P(E,t)$ is the probability of having energy $E$ at time $t$, and 
\begin{equation}
  T(E,E') = {1 \over n(E') } \sum_{H(\sigma) = E} \>\sum_{H(\sigma') = E'}
\!\!\! \Gamma(\sigma, \sigma'),
\end{equation} 
where $n(E)$ is the degeneracy of the states.  We can not derive
Eq.~(\ref{Teq}) from Eq.~(\ref{Meq}) in general, the ``derivation'' is
valid only at equilibrium when $P(E) = \sum_{H(\sigma) = E} P(\sigma)
= n(E) \exp(-E/k_BT)$.

The transition matrix $T(E,E')$ has some general properties: (1) The
matrix is banded alone diagonal.  (2) The column sum is zero,
$\sum_E T(E,E') = 0$, due to the conservation of total probability.
(3) $\sum_{E'} T(E,E') P_{eq}(E') = 0$, due to existence of
equilibrium distribution.  (4) The transition rate satisfies detailed
balance conditions, $T(E',E) P_{eq}(E) = T(E,E') P_{eq}(E')$.

The transition matrix Monte Carlo dynamics \cite{Wang-Tay-Swendsen}
has the following interesting features: (1) It is a constrained
random walk in energy space.  (2) The transition rates are derived
from single-spin-flip dynamics.  (3) It has a fast dynamics, $\tau
\propto c$, and (4) it suggests a different histogram reweighting
method.

The artificial dynamics described by Eq.~(\ref{Teq})
can be implemented on computer in at least
two different ways, we'll call them algorithm A and B.

\medskip
{\noindent\bf Algorithm A}
\nobreak
\begin{enumerate}
  \item Do sufficient number of constant energy (microcanonical) Monte
Carlo steps, so that the final configuration is totally uncorrelated
with the initial configuration.  This step is equivalent to pick a
state $\sigma$ at random from all states with energy $E$.

   \item Do one single-spin-flip canonical Monte Carlo move.
\end{enumerate}
Clearly, this algorithm is not very efficient computationally, due to
step 1.  However, it will be helpful in understanding the dynamics.

\medskip
{\noindent\bf Algorithm B}
\nobreak
\begin{itemize}
\item A direct implementation of Eq.~(\ref{Teq}), i.e., a random
walk in energy with a transition rate $T(E,E')$.
\end{itemize}

Then in algorithm B, we need to know $T(E,E')$ explicitly, this can be
done numerically by Monte Carlo sampling, from
\begin{equation}
  T(E+\Delta E,E) = w(\Delta E) 
\bigl\langle N(\sigma, \Delta E) \bigr\rangle_E, \qquad
\Delta E \neq 0,
\end{equation}
and $w(\Delta E) = {1\over 2} \bigl( 1 - \tanh(\Delta
E/(2k_BT)\bigr)$.  $N(\sigma, \Delta E)$ is the number of cases that
energy is changed by $\Delta E$ from $E$ for the $N$ possible
single-spin flips.

Note that computation of $\bigl\langle N(\sigma, \Delta E)
\bigr\rangle_E$ can be done with any sampling technique
which ensures equal probability for equal energy.  We
use canonical simulations at selected
temperatures to compute the microcanonical average 
$\bigl\langle N(\sigma, \Delta E) \bigr\rangle_E$
so that the total histogram is roughly flat.
Alternative sampling methods are given in ref.~\cite{Isbroad}.
The transition
matrix can be formed with any temperature.  The equilibrium
distribution and thus the density of states $n(E) = P_{eq}(E)
\exp(E/k_BT)$ is obtained by solving
\begin{equation}
   \sum_{E'} T(E,E') P_{eq}(E') = 0,
\end{equation}
or by solving a set of detailed balance conditions. 
The above scheme is similar in spirit to the histogram method of
Ferrenberg and Swendsen \cite{Ferrenberg-Swendsen}, and the method has
a close connection with, but different from the broad histogram of
Oliveira et al \cite{Oliveira}.  In Fig.~1, we show the determination
of two-dimensional Ising model average energy on a $64 \times 64$
lattice.  The errors are very small on a whole temperature range.

We have more or less a complete understanding of the transition matrix
Monte Carlo dynamics through exact results in limiting cases.  The
transition matrix $T(E,E')$ can be computed exactly in one-dimensional
chain of length $L$ (with periodic boundary condition), by some
combinatorial consideration, as
\begin{eqnarray}
    T_{k, k+1} & = &  { (k+1)(2k+1) \over L - 1} (1+\gamma), \\
    T_{k+1,k} & =  & { (L-2k)(L-2k-1) \over 2 (L-1) } (1-\gamma),
\end{eqnarray}
where $\gamma = \tanh(2J/k_BT)$.
The diagonal terms are computed from the relation
\begin{equation}
    T_{k-1,k} + T_{k,k} + T_{k+1,k} = 0,
\end{equation}
and the rest of the elements $T_{k,k'} = 0$ if $|k-k'| > 1$.  The
integer $k=0, 1, 2, \ldots, \lfloor L/2 \rfloor$ is related to energy
by $E/J= -L + 4k$.  While the eigen spectrum at temperature $T=0$ can
be computed exactly as $\lambda_k = - 2(k+1)(2k+1)/(L-1)$, the
eigenvalues at $T>0$ is obtained only numerically.  The most important
feature is that $\tau \propto L$, given an unusual dynamical critical
exponent of $z=1$ in one dimension.

The dynamics in any dimensions in the large size limit
\cite{VanKampen} obeys a linear Fokker-Planck equation:
\begin{equation}
   {\partial P(x,t') \over \partial t'} = 
{ \partial \over \partial x } \left( {\partial P(x,t') \over \partial x }
+ x P(x,t') \right), \label{diffusionEq}
\end{equation}
where $t'$ and $x$ are properly scaled time and energy.   
\begin{equation} 
  x = { E - u_0 N \over ( N c' )^{1/2} },  \quad u_0 N = \bar E,
\end{equation}
and $t' = b t$ with
\begin{equation}
  b = \lim_{N\to\infty} {1\over 2 c' N} \sum_{E'} T(\bar E,E') (E'-\bar E)^2,
\label{beq}
\end{equation}
where $u_0$ is the average energy per spin and $c'=k_BT^2c$ is the
reduced specific heat per spin.  The major consequence of this result
is that the relaxation times are $\tau_n = a c'/n$, $n=1,2,3,\cdots$,
with some constant $a$.

The exact results can be interpreted with intuitive pictures.  First,
we consider the result of $\tau \propto L$ in one dimension as $T \to
0$.  At sufficiently low temperatures with a correlation length $\xi$
comparable with the system size $L$, only the ground state (all spins
up or down) and the first excited states (with a kink pair) are
important.  Let's consider the time scale for $E_0 \to E_1$. A spin
with opposite sign is created with probability $\exp(-4K)$ from
Boltzmann weight, where $K=J/(k_BT)$, in each of the canonical move.
Thus
\begin{equation}
   \tau \propto { \exp(4K) \over L } \propto {\xi^2 \over L} \propto L.
\end{equation}
where $K$ is chosen such that there is about one kink pair, so
that $\xi \sim \exp(2K) \sim L$.

Similarly, the result of $\tau \propto c$ can be obtained by the
following argument.  The transition matrix Monte Carlo is a random
walk constrained in the range $\delta E$, due to the gaussian
distribution nature of the equilibrium distribution $P_{eq}(E)$.  The
width of this distribution is related to the specific heat by $\delta
E^2 = c N k_BT^2$.  Each walk changes $E$ by $O(1)$. To change $E$ by
$\delta E$, we need $\delta E^2$ moves, invoking the theory on random
walks.  In units of transition matrix Monte Carlo steps,
\begin{equation}
  \tau \approx a { \delta E^2 \over N} \propto  c.
\end{equation}

Part of the work presented in this talk is in collaboration with Tay
Tien Kiat and Robert~H.~Swendsen.  This work is supported in part by a
NUS Faculty Research Grant PR950601.

\begin{figure}
\input epsf.tex
\epsfxsize=\hsize\epsffile{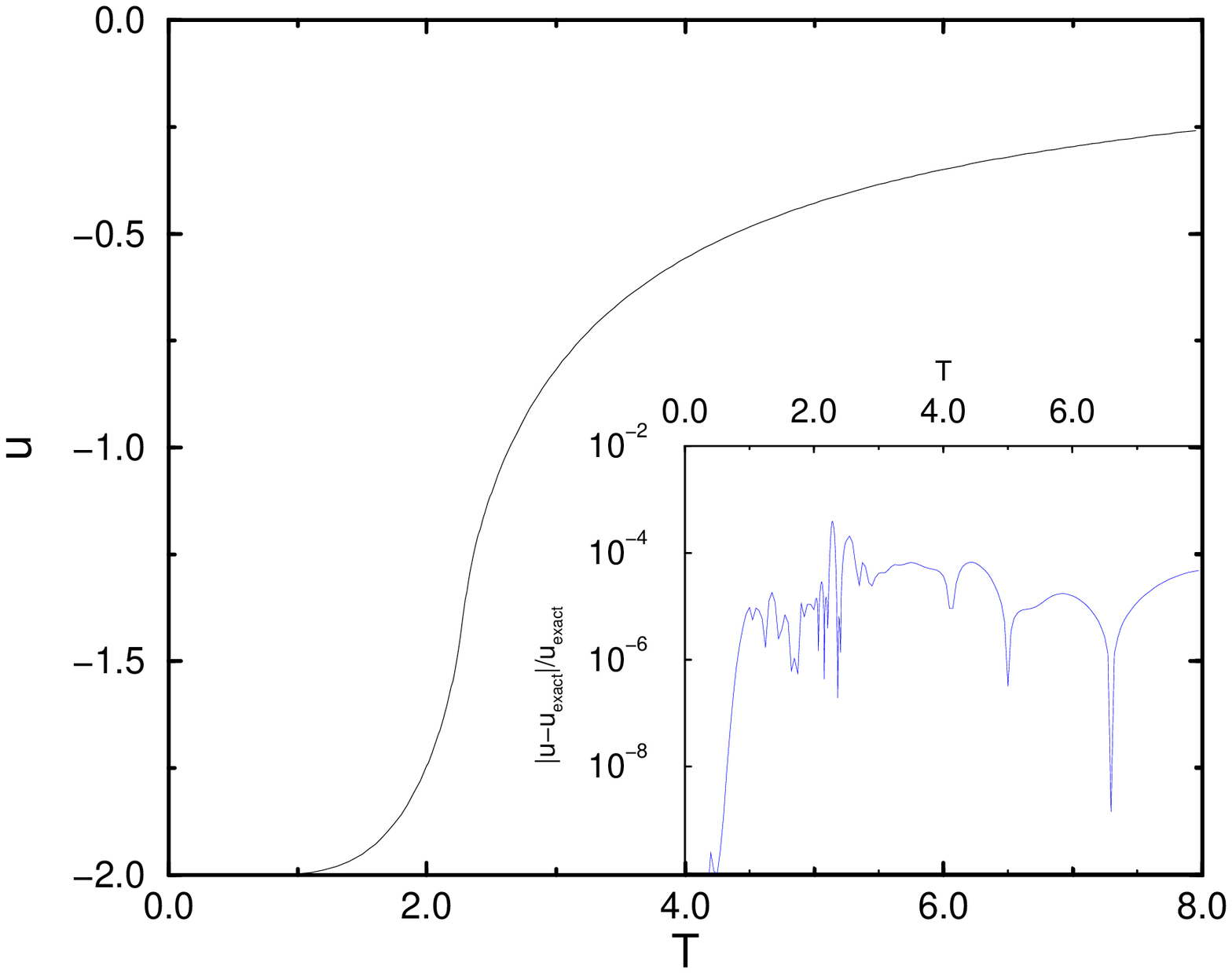}
\caption[fig1]{The average energy per spin of the two-dimensional Ising model 
on a $64^2$ lattice by the transition matrix Monte Carlo reweighting
method.  The insert shows the relative error with respect to the exact
result (obtained numerically based on \cite{exact}).  The canonical simulations
are performed at 25 temperatures, each with $10^6$ Monte Carlo steps with a
single-spin-flip dynamics.  }
\end{figure}
\end{document}